\RequirePackage{amsmath}
\documentclass[runningheads]{llncs}
\usepackage{graphicx}
\usepackage[utf8]{inputenc} 
\usepackage[T1]{fontenc}   
\usepackage{url}           
\usepackage{booktabs}     
\usepackage{amsfonts}
\usepackage{nicefrac}      
\usepackage{microtype}      
\usepackage{enumerate}
\usepackage{graphicx}
\usepackage[caption=false]{subfig}
\usepackage{algorithmic}
\usepackage{algorithm2e}

\begin{document}
\title{
Estimating Uncertainty and Interpretability in Deep Learning for Coronavirus (COVID-19) Detection
}

\titlerunning{Estimating uncertainty in Deep Learning for COVID-19 Detection}
%
%

\author{Biraja Ghoshal\inst{1}
\and
Allan Tucker\inst{1}
}

\authorrunning{B. Ghoshal et al.}
%
\institute{Brunel University London, Uxbridge, UB8 3PH, United Kingdom 
\\
\email{biraja.ghoshal@brunel.ac.uk}\\
\url{https://www.brunel.ac.uk/computer-science}}
\maketitle              

\begin{abstract}
Deep Learning has achieved state of the art performance in medical imaging. However, these methods for disease detection focus exclusively on improving the accuracy of classification or predictions without quantifying uncertainty in a decision. Knowing how much confidence there is in a computer-based medical diagnosis is essential for gaining clinicians’ trust in the technology and therefore improve treatment. Today, the 2019 Coronavirus (COVID-19) infections are a major healthcare challenge around the world. Detecting COVID-19 in X-ray images is crucial for diagnosis, assessment and treatment. However, diagnostic uncertainty in a report is a challenging yet inevitable task for radiologists. In this paper, we investigate how Dropweights based Bayesian Convolutional Neural Networks (BCNN) can estimate uncertainty in Deep Learning solutions to improve the diagnostic performance of the human-machine combination using publicly available COVID-19 chest X-ray dataset and show that the uncertainty in prediction is strongly correlated with the accuracy of the prediction. We believe that the availability of uncertainty-aware deep learning will enable a wider adoption of Artificial Intelligence (AI) in a clinical setting. \\

  
\end{abstract}

\section{Introduction}

In recent years, Deep Learning has achieved state of the art performance, similar to that of
human experts in solving classification tasks in computer vision from lung disease classification, metastasis detection for breast cancer, skin lesion classification, identifying diabetic retinopathy, attention deficit hyperactivity disorder (ADHD), Alzheimer's disease and improving reconstruction for MRI, PET/CT imaging. However, despite the promising results, deep learning for classification tasks
lacks the ability to say “I don’t know” in an ambiguous or unknown case. Hence, it is critical to estimate uncertainty in medical imaging as an additional insight to point predictions to improve the reliability in making decisions.

Dealing with Coronavirus (COVID-19) is one of the major healthcare challenges around the world today. COVID-19 represents a new strain of Coronavirus and presumably representing a mutation of other Coronaviruses \cite{shan+2020lung}. 

The existing infrastructure (e.g. limited image data sources with expert labelled data set) for the detection of COVID-19 positive patients is insufficient and manual detection is time-consuming. With the increase in  global incidences, it is expected that a Deep learning based solution will soon be developed and combined with clinical practices to provide cost-effective, accurate and easily performed automated detection of COVID-19 to aid the screening process. 

However, despite remarkable performance, deep learning models tend to make overconfident predictions. Our objective is not to achieve state-of-the-art performance, but rather to evaluate the usefulness of estimating uncertainty approximating Bayesian Convolutional Neural Networks (BCNN) with Dropweights to improve the diagnostic performance of combined human-machine \cite{Ghoshal2020EstimatingUncertainty,ghoshal2019journal}. This is crucial in differentiating COVID-19 patients from those without the disease,
where the cost of an error is very high. Thus, in order to avoid COVID-19 misdiagnoses \cite {li2020coronavirus}, it is necessary to estimate uncertainty in a model's predictions. 

In this paper, we investigate how  Monte-Carlo Dropweights (MC Dropweights) 
Bayesian convolutional neural networks can estimate uncertainty in Deep Learning to improve the diagnostic performance of human-machine decisions, using publicly available COVID-19 chest X-ray datasets, and show that the estimated uncertainty in prediction has a strong correlation with classification accuracy, thus enabling the identification of false predictions or unknown cases. 

\section{Related Research}
Estimating uncertainty in deep neural networks is a challenging and unsolved problem. There are many measures to estimate uncertainty such as softmax variance, expected entropy, mutual information, predictive entropy and averaging predictions over multiple models. 

Bayesian Neural Networks (BNN) provides a natural framework for modelling uncertainty \cite{blundell2015weight}. However, BNN methods are intractable in computing the posterior of a network’s parameters. The most used approach to estimate uncertainty in deep learning try to place distributions over each of the network's weight parameters \cite{blundell2015weight} of a model. 

There are many methods proposed for quantifying uncertainty or confidence estimates approximated by Monte-Carlo Dropout, including Laplace approximation, Markov chain Monte Carlo (MCMC) methods, stochastic gradient MCMC variants such as Langevin Dynamics, Hamiltonian methods, including Multiplicative Normalizing Flows, Stochastic Batch Normalization, Maximum Softmax Probability, Heteroscedastic Classifier, and Learned Confidence Estimates including Deep Ensembles \cite{YarinGal2016Thesis}.

\section{Approximate Bayesian Convolutional Neural Networks (BCNN) and Model Uncertainty}

Given dataset \(X=\left\{x_{1}, x_{2} \ldots x_{N}\right\}\) and the corresponding labels \(Y=\left\{y_{1}, y_{2} \ldots y_{N}\right\}\) where \(X \in R^{d}\) is a d-dimensional input vector and \(Y \in\{1 \ldots C\}\) with \(\mathrm{y}_{\mathrm{i}} \in\{1 \ldots \mathrm{C}\}\), given C class label, a set of independent and identically distributed (i.i.d.) training samples size \(N\)\(\left\{x_{i}, y_{i}\right\}\)  for \(i=1\) to \(N\), the objective is to find a function \(f : X \rightarrow Y\) using weights of neural net parameters \(w\) as close as possible to the original function that has generated the outputs \(\hat{Y}\). The principled predictive distribution of an unknown label \(\hat{y}\) of a test input data \(\hat{x}\) by marginalizing the parameters:

\begin{equation}
p(\hat{y}|\hat{x}, \mathcal{X, Y}) = \int P(\hat{y} | \hat{x}, {w}) P({w}| X, Y, \hat{x}) d{w}
\end{equation}

Unfortunately, finding the posterior distribution \(p({w}|{X, Y})\) is often computationally intractable. Recently, Gal \cite{YarinGal2016Thesis} proved that a gradient-based optimization procedure
on the dropout neural network is equivalent to a specific variational approximation on a Bayesian neural network. Following Gal \cite{YarinGal2016Thesis}, Ghoshal et al. \cite{19Biraja} also showed similar results for neural networks with MC-Dropweights. 
The model uncertainty is approximated by averaging stochastic feed forward Monte Carlo (MC) sampling during inference. At test time, the unseen samples are passed through the network before the Softmax predictions are analyzed.

Practically, the expectation of \(\hat{y}\) is called the predictive mean of the model. The predictive mean \(\mu_{pred}\) over the MC iterations is then used as the final prediction on the test sample:
\begin{equation}
\mu_{pred} \approx \frac{1}{T} \sum_{t=1}^{T} p(\hat{y}|\hat{x}, \mathcal{X, Y})
\end{equation}

For each test sample \(\hat{x}\), the class with the largest predictive mean \(\mu_{pred}\) is selected as the output prediction and the variance is the predictive uncertainty.

\subsection{Uncertainty Estimation in Classification}

In order for COVID-19 detection to be meaningful, tolerance must typically be much tighter. Based on the input X-ray image, a network can be certain with high or low confidence about its decision, indicated by the predictive posterior distribution.

However predictive uncertainty in deep learning actually results from two separate forms of uncertainty \cite{depeweg2017decomposition}:
\begin{enumerate}
\item	Epistemic uncertainty or Model uncertainty accounts for uncertainty in the model parameters as it does not take all of the aspects of the data into account or the lack of training data. Epistemic uncertainty associated with the model reduces as the training data size increases. 
\item	Aleatoric uncertainty accounts for noise inherent in the observations due to class overlap, label noise, homoscedastic and heteroscedastic noise, which cannot be reduced even if more data were to be collected. In X-ray imaging, this can be caused by sensor noise due to random distribution of photons during scan acquisition.
\end{enumerate}

Traditionally, it has been difficult to implement model validation under epistemic uncertainty. Thus, we estimate epistemic uncertainty to obtain model uncertainty in deep learning prediction for chest radiograph diagnosis for COVID-19. 
One of the measure of model uncertainty is predictive entropy \(H\) of the predictive distribution: 
\begin{equation}
H(\hat{y}|\hat{x}, \mathcal{X, Y}) =  
-\sum_{C}p(\hat{y}=c| \hat{x},\mathcal{X, Y})\log p(\hat{y}=c| \hat{x},\mathcal{X, Y})
\end{equation}

where C ranges over all class labels. In general, the range of the obtained uncertainty values depend on datasets, network architectures, number of MC sampling, etc. Therefore, we normalise estimated uncertainty to report our results and facilitate the comparison across various sets and configurations.

Our analysis involved a comparison of two variational-dropweights based uncertainty measures, Predictive Entropy (PH) and Bayesian Active Learning by Disagreement (BALD)\cite{houlsby2014efficient,smith2018understanding}, in their application to COVID-19 image classification. 

The second uncertainty measure, Bayesian Active Learning by Disagreement (BALD), is based on mutual information that maximise the mutual information between model posterior density function and predictions density function approximated at as the difference between the entropy of the predictive distribution and the mean entropy of predictions across samples: 
\begin{equation}
MI\left[\hat{y}_{i}, w | \hat{x}_{i}, \mathrm{X}, \mathrm{Y}\right] \approx H\left[\hat{y}_{i} | \hat{x}_{i}, {X}, {Y}\right]-E\left[H\left[\hat{y}_{i} | \hat{x}_{i} w \right]\right]
\end{equation}
, with \(w\) the model parameters.

Test points that maximise mutual information are points over which the model is uncertain on average, but there are model parameters that produce erroneous predictions with a high confidence. This is equivalent to points with high variance in the input to the softmax layer (the logits). Thus, each stochastic forward pass through the model would have the highest probability assigned to a different class. It is expected from BALD measures epistemic uncertainty of the model, so it would not return a high value if there is aleoratic uncertainty present.

\subsection{Relationship between the Accuracy and Uncertainty}

The true error is the difference between estimated values and actual values. In order to assess the quality of predictive uncertainty, we leveraged Spearman’s correlation coefficient between Predictive Entropy (PH) and Bayesian Active Learning by Disagreement (BALD). We quantified the predictive accuracy by 1-Wasserstein distance (WD) to measure how much the estimated uncertainty correlates with the true errors \cite{arjovsky2017wasserstein,laves2019quantifying}.
The Wasserstein distance for the real data distribution \(P_r\) and the generated data distribution \(P_g\) is mathematically defined as the greatest lower bound (infimum) for any transport plan (i.e. the cost for the cheapest plan):
\begin{equation}
W(P_r, P_g) = \inf_{\gamma \sim \Pi(P_r, P_g)} \mathbb{E}_{(x, y) \sim \gamma}[\| x-y \|]
\end{equation}, \(\Pi(P_r, P_g)\) is the set of all possible joint probability distributions \(\gamma(x, y)\) whose marginals are respectively \(P_r\) and \(P_g\). However, the equation (5) for the Wasserstein distance is intractable. Using the Kantorovich-Rubinstein duality, \cite{arjovsky2017wasserstein} simplified the calculation to

\begin{equation}
W(P_r, P_g) = \frac{1}{K} \sup_{\| f \|_L \leq K} \mathbb{E}_{x \sim P_r}[f(x)] - \mathbb{E}_{x \sim P_g}[f(x)]
\end{equation}, where sup (supremum) is the opposite of inf (infimum); sup is the least upper bound and \(f\) is a 1-Lipschitz continuous functions \(\{ f_w \}_{w \in W}\), parameterized by \(w\) and the K-Lipschitz constraint \(\lvert f(x_1) - f(x_2) \rvert \leq K \lvert x_1 - x_2 \rvert\). The error function can be configured as measuring the 1 - Wasserstein distance between \(P_r\) and \(P_g\).
\begin{equation}
E(P_r, P_g) = W(P_r, P_g) = \max_{w \in W} \mathbb{E}_{x \sim P_r}[f_w(x)] - \mathbb{E}_{z \sim P_r(z)}[f_w(g_\theta(z))]
\end{equation}

The advantage of Wasserstein distance (WD) is that it can reflect the distance of two non-overlapping or little overlapping distributions.

\section{Dataset}
Radiologists frequently use X-ray images to detect lung inflammation, enlarged lymph nodes or pneumonia. Once the COVID-19 virus is inside the body, it begins infecting epithelial cells lining the lung. We can use X-rays to analyse the health of a patient’s lungs. Analysis of X-ray requires an expert and takes significant time.

\subsection {Data Preparation}
We have selected 68 Posterior-Anterior (PA) X-ray images of lungs with COVID-19 cases from Dr. Joseph Cohen’s Github repository \cite{20Cohen}. We augmented the dataset with Kaggle’s Chest X-Ray Images (Pneumonia) from healthy patients, a total of 5941 PA chest radiography images across 4 classes (Normal: 1583, Bacterial Pneumonia: 2786, non-COVID-19 Viral Pneumonia: 1504, and COVID-19: 68).

\newpage
\section{Experiment}

Instead of training a very deep model from scratch on a small dataset, we decided to run this experiment in a transfer learning setting, where we used a pre-trained ResNet50V2 model \cite{he2016identity} and acquired data only to fine-tune the original model. This is very suitable when the data is abound for an auxiliary domain, but very limited labelled data is available for the domain of experiment. We introduced fully connected layers on top of the ResNet50V2 convolutional base. Dropweights followed by a softmax activated layer is applied to the network as an approximation to the Gaussian Process (GP) and to cast it as an approximate Bayesian inference, in the fully connected layer to estimate meaningful model uncertainty. The softmax layer outputs the probability distribution over each possible class label. 

We resized all images to 224 x 224 pixels (using a bicubic interpolation over 4 x 4 pixel neighbourhood). The images were standardised using the mean and standard deviation values of the X-ray dataset.
We split the whole dataset into 80\% - 20\% between training and testing sets. Real-time data augmentation was also applied, leveraging Keras ImageDataGenerator during training, to prevent overfitting and enhance the learning capability of the model. Training images were ZCA whitened, rotated 20 degree, randomly flipped horizontally and vertically, scaled outward and inward, shifted, and sheared. The Adam optimiser was used with a learning rate of 1e-5 with decay factor of 0.2. All our experiments were run for 25 epochs and batch size was set to 8. 
Dropweights with rates of \{0.1, 0.3, and 0.5\} were added to a fully-connected layer. We monitored the validation accuracy after every epoch and saved the model with the best accuracy on the validation dataset. During test time, Dropweights were active and Monte Carlo sampling was performed by feeding the input image with MC-samples \{10, 25 and 50\} through the Bayesian Deep Residual Neural Networks.

\subsection{Asymmetric Cost Function}
The cost of falsely diagnosing of COVID-19, when a patient does not have it (i.e. a false positive result) may be much lower than not detecting a COVID-19 case, when it is present (i.e. a false negative result). Our goal is to avoid any false negative detection, even if it means that some false positives are incurred.

The asymmetric cost of making mistakes is captured by a utility function \cite{cobb2018loss} such as class weights which dictates optimal predictions while to approximate the true posterior over the weights. In order to address this asymmetric cost of making mistakes, we have defined utility function \((\alpha)\) in maximising the expected utility. So the weighted cross-entropy loss function is defined as:

\begin{equation}
 L \approx \frac{1}{C} \sum_{c=1}^{C} \alpha_{c} * p(\hat{y} = c_{i} | w, \hat{x})
\end{equation}
, where \(\alpha_{c}\) is the corresponding weight for each class, c, in the cross-entropy loss. The above equation dictates optimal predictions to approximate the true posterior over the weights. The highest weight \{Normal: 2; Bacterial: 2; Viral: 1; COVID-19: 50\} is assigned for an image which is misdiagnosed as being non-COVID-19 infected when ground truth is true with low uncertainty.

\section{Results and Discussions}

\subsection{Uncertainty-Aware Prediction Performance of the Bayesian Models}

Most of COVID-19 cases' chest X-rays show bilateral pulmonary infiltrates with distinctive appearances. The below Figure 1 shows the distribution of predictive uncertainty values for all test X-Ray images, grouped by correct (in green) and erroneous (in red) predictions. The class with the highest softmax output for predictive distribution mean is considered as the prediction and the predictive entropy of the output distributions (measured as in Equation (3)) as the estimated epistemic uncertainty. Based on the input image, a network can be certain with high or low confidence about its decision, indicated by the predictive posterior distribution. The wider the output posterior distributions, the less confident is the model in it's prediction. This is because the uncertainty in weight space captured by the posterior is incorporated into the predictive uncertainty, giving us a way to model to say “I don’t know”.

\begin{figure} [!h]
    \centering
    \includegraphics[width=10cm]{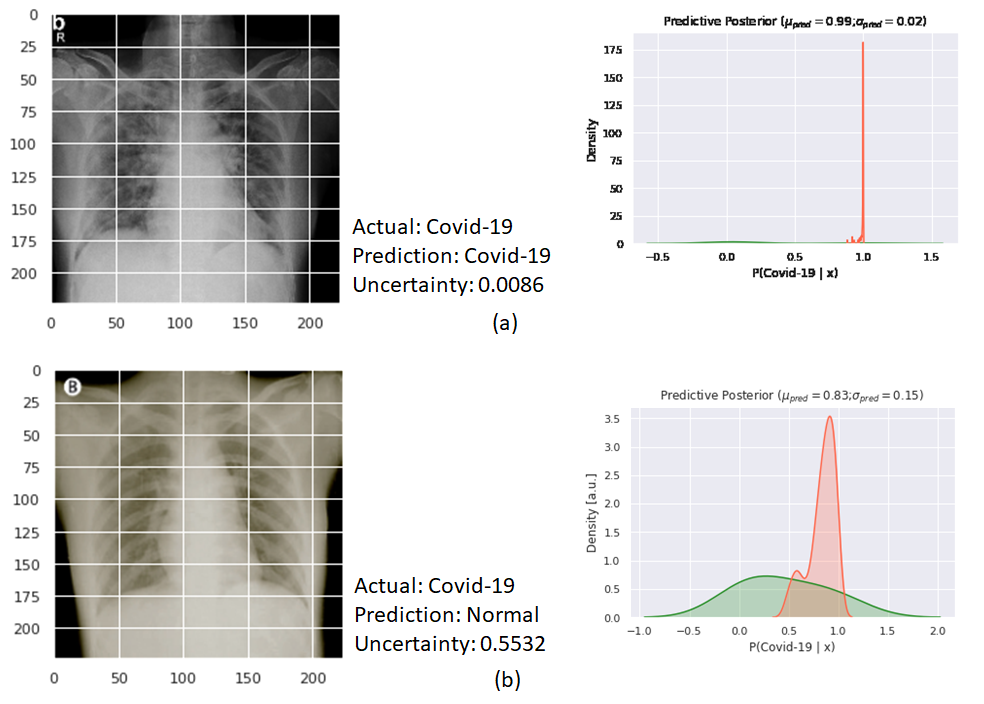}
    \caption{Example input images with uncertainty and the corresponding predictive distributions generated by Bayesian DNN. Figure 1(a) shows a correctly classified image where the model is highly certain about its prediction (PH=0.0086). Whereas, figure 1(b) shows a miss-classified image where the model is uncertain (PH=0.55332) and wider posterior distributions.}
\end{figure}

\newpage
\subsection{Bayesian Models Performance}
On average, Bayesian ResNet50V2 model based inference improves the detection accuracy of the standard ResNet50V2 model in our sample dataset based solely on X-ray images. Figure 2 confusion matrix summarizes the prediction accuracy of our implemented models.

\begin{figure} [!ht]
    \centering
    \includegraphics[width=12cm]{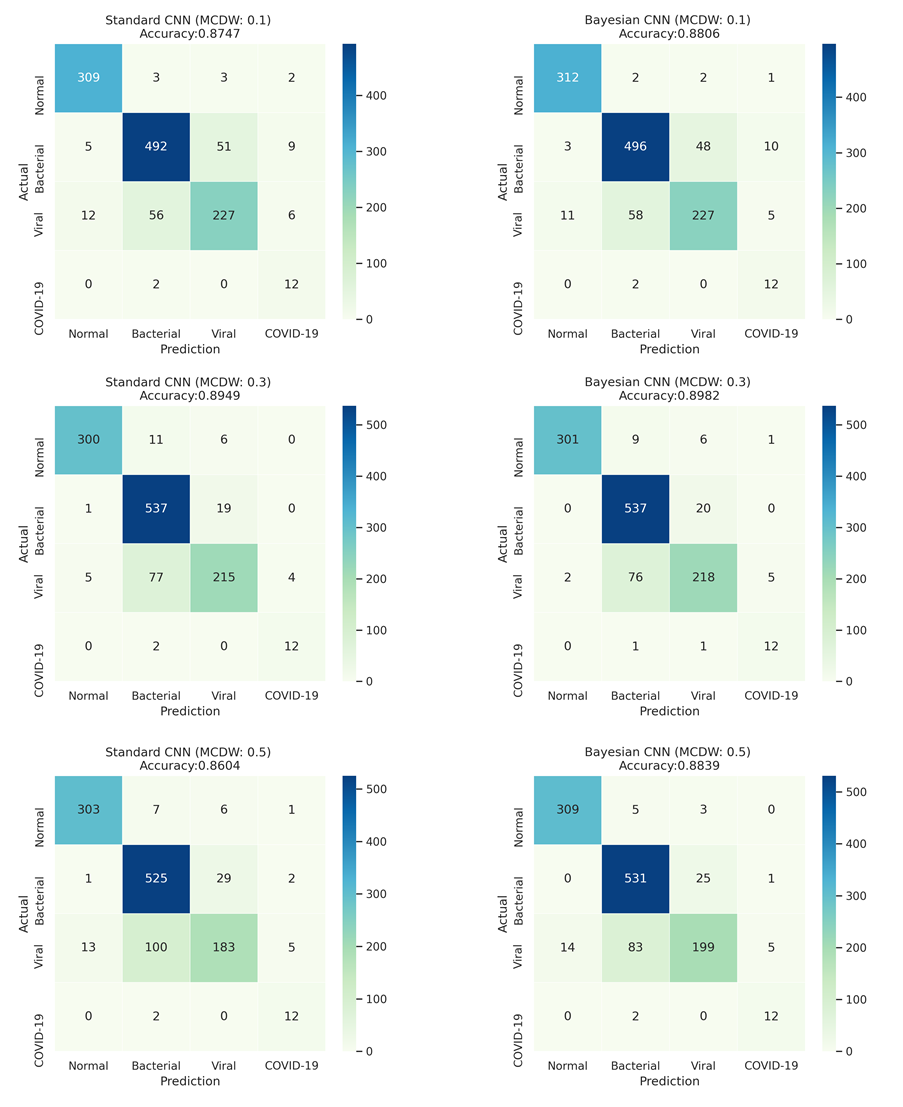}
    \caption{Confusion Matrix }
\end{figure}

\subsection{Bayesian Model Uncertainty}

We measured the epistemic uncertainty associated with the predictive probabilities of the deep learning model by keeping dropweights on during test time. Figure 3 below shows Kernel Density Estimation with a Gaussian kernel is used to plot the output posterior distributions for all X-Rays test images, grouped by correct and erroneous predictions with variation of dropwights rate \(p\) for 50 MC samples of stochastic feed forward. The table below shows the effect of variation of the dropweights rate, \(p\), to the uncertainty measures.

\begin{figure} [!ht]
    \centering
    \includegraphics[width=\linewidth]{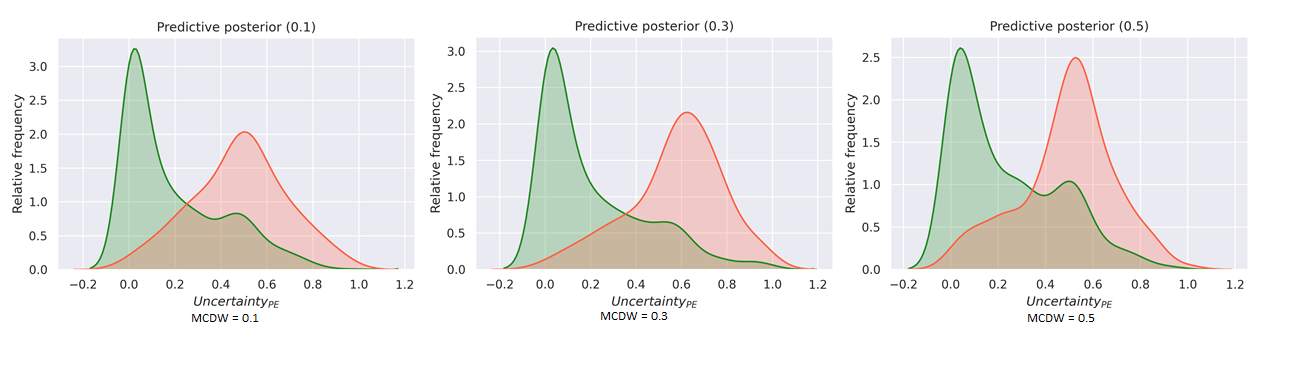}
    \caption{Distribution of estimated predictive uncertainty for all test samples grouped by correct and erroneous predictions.}
\end{figure}

It shows that the estimated uncertainty is higher for erroneous predictions. Therefore, uncertainty information provides as an additional insight to point prediction to refer the uncertain images to radiologists for further investigation \cite{laves2019quantifying}, which improves the overall prediction performance. 

\begin{figure} [!ht]
    \centering
    \includegraphics[width=12cm]{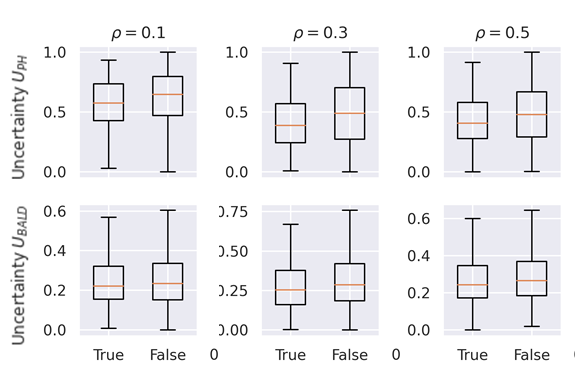}
    \caption{Quality of Uncertainty measure in Covid-19 Chest X-Ray Detection \cite{laves2019quantifying} }
\end{figure}

Figure 4 shows the effect of variation of the Dropweights rate  \(p\) to the uncertainty measures (PH and BALD). The  results suggest, that predictive entropy as a measure of uncertainty is a better measure for uncertainty and should be considered over BALD. Regardless of values for the number of MC samples and Dropweights rate, we can observe a higher uncertainty for incorrect classification. MC dropweights for uncertainty estimation can usually be used in every image classifier to improve prediction accuracy of man–machine combination via uncertainty-aware referral with the additional computational load cost of performing multiple forward passes.

\newpage
\subsection{The relation between uncertainty and predictive accuracy}

The table 1 in below shows that there is strong correlation between predictive entropy and the prediction error.

\begin{table}[!h]
\centering
\begin{tabular}{|l|l|l|}
\hline
Spearmans's Correlation & Predictive Entropy & BALD   \\ \hline
Dropweights Rate:0.5 & 0.9951 & 0.8754 \\ \hline
Dropweights Rate:0.3 & 0.9968 & 0.8873 \\ \hline
Dropweights Rate:0.1 & 0.9952 & 0.8980 \\ \hline
\end{tabular}
\end{table}

The figure 5 below shows the correlation between estimated uncertainty from PH and BALD and the
error of prediction with variation of the Dropweights rate \(P\). The above results show strong correlation with \(\rho\) = 0.99 between entropy of the probabilities as a measure of the epistemic uncertainty and prediction errors.

\begin{figure} [!ht]
    \centering
    \includegraphics[width=\linewidth]{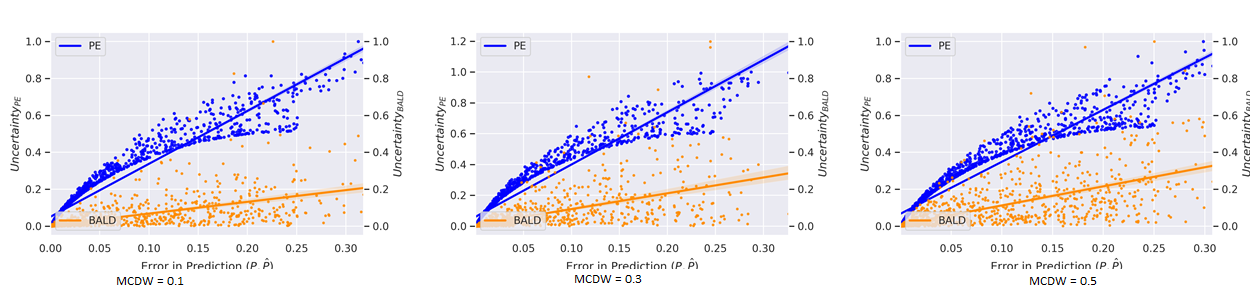}
    \caption{Correlation between estimated predictive entropy as a measure of Uncertainty and Accuracy in prediction \cite{laves2019quantifying}}
\end{figure}

Our experiments show that the prediction uncertainty correlates with accuracy, thus enabling the  identification of false predictions or unknown cases.

\newpage
\subsection{Performance improvement via Uncertainty-Aware COVID-19 Classification and Referral}

We performed predictions for all COVID-19 test images and sorted the predictions by their associated predictive uncertainty (PH). We then referred predictions based on the various levels of  uncertainty for further diagnosis and measured the accuracy of the predictions (threshold at 0.5) for the remaining cases. We observed in the figure 6, the prediction accuracy increases with the fraction of referred images. Note that only non-referred images are considered to compute predictive accuracy. We have also observed the same behaviour in prediction accuracy for increasing levels of model uncertainty.

\begin{figure} [!ht]
    \centering
    \includegraphics[width=\linewidth]{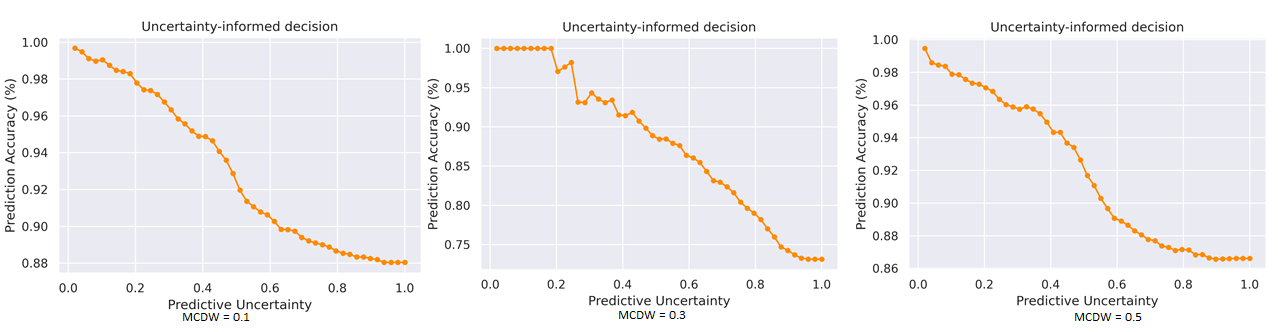}
    \caption{The classification accuracy as a function of the tolerated normalized model uncertainty }
\end{figure}

\begin{figure} [!ht]
    \centering
    \includegraphics[width=\linewidth]{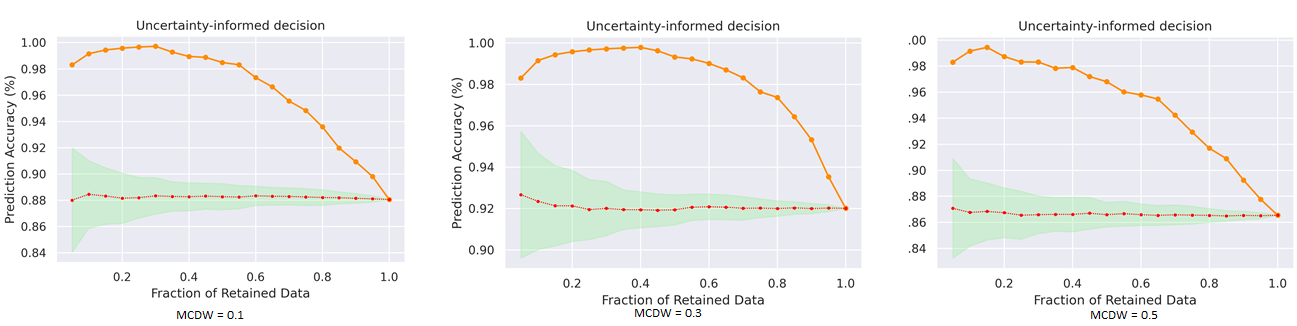}
    \caption{The classification accuracy as a function of the retained data}
\end{figure}

Simulating a control experiment, we compared with randomly selected images, that is without using uncertainty in prediction (figure 7).

For a beginner radiologist performance (i.e. 60\% prediction accuracy), solely relying on deep learning models will result in a more accurate prediction on overall diagnosis. However, for an experienced radiologist (i.e. 80\% accuracy), the combined performance reaches almost 90\% when rejecting either almost 40\% of the most uncertain samples or samples with \(H_{norm} >=0.4\).
For less than 2\% decisions referred for further inspections, there is a 95\% confidence interval of the two non-overlapping scenarios. Hence,  estimated uncertainty provides as an additional insight to point prediction performance to improve the reliability of the automated system. 

\section{Visualizing Uncertainty and Interpretability}
Deep learning models often been accused of being "black boxes", so they need to be precise, interpretable and the uncertainty in
predictions must be well understood.
Reliable estimated uncertainty alongside the visualisation of distinct features, as an additional insight to point
prediction, will improve the ease of understanding in deep learning, resulting in a more informed decision-making process.
We qualitatively compare in figure 8, the saliency maps \cite{adebayo2018sanity} produced by various state-of-the-art methods e.g.Class Activation Map (CAM), Guided Backpropagation and Guided Gradient CAM and  Gradients.

\begin{figure} [!ht]
    \centering
    \includegraphics[width=\linewidth]{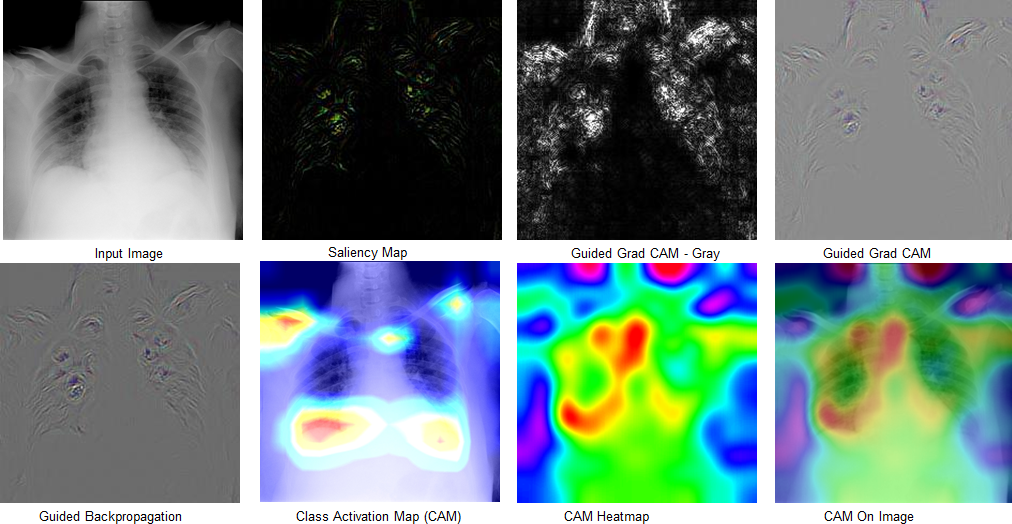}
    \caption{Saliency Map using various methods }
\end{figure}

\section{Conclusion and Future work}

In this work, Bayesian Deep Learning  classifier has been trained using transfer learning method on COVID-19 X-Ray images to estimate model uncertainty. Our experiment has shown a strong correlation between model uncertainty and accuracy of prediction. The estimated uncertainty in deep learning yields more reliable prediction, which can alert radiologists on false predictions, which will increase the acceptance of deep learning into clinical practice in disease detection.

With this Bayesian Deep Learning based classification, studies correlating with multi "omics" dataset \cite{gozes2020rapid}, and treatment responses could further reveal insights about imaging markers and findings towards improved diagnosis and treatment for Covid-19.
%
%
\bibliographystyle{splncs04}
\bibliography{Covid}
\end{document}